\documentstyle[psfig]{lamuphys}
\makeatletter
\let\chapter\hid@chapter
\makeatother
\begin{document}
\pagenumbering{arabic}

\title{Tidal thickening of galaxy disks}

\author{Vladimir\,Reshetnikov\inst{1,2}, Francoise\,Combes\inst{2}}

\institute{Astronomical Institute of St.Petersburg State University, 
St.Petersburg, Russia 
\and
DEMIRM, Observatoire de Paris, 61 Av. de l'Observatoire, F-75014
Paris, France}

\maketitle

\begin{abstract}
We have studied a sample of 24 edge-on interacting galaxies and
compared them to edge-on isolated galaxies, to investigate the effects
of tidal interaction on disk thickening. We found that 
the ratio $h/z_0$ of the
radial exponential scalelength $h$ to the scale height $z_0$
is about twice smaller for interacting galaxies. This is found to be
due both to a thickening of the plane, and to a radial stripping
or shrinking of the stellar disk. 
If we believe that any galaxy experienced a tidal interaction in 
the past, we must conclude that continuous gas accretion and
subsequent star formation can bring back the ratio $h/z_0$ to
higher values, in a time scale of 1 Gyr.
\end{abstract}

\section{Motivation}

Galaxy disks are very sensitive to tidal interactions, from the formation 
of tidal tails and bridges up to the complete disruption of initial disks
in mergers. Even non-merging 
interactions or minor mergers can thicken and destroy a stellar disk, and 
this
has been advanced as an argument against frequent interactions in
a galaxy life, or formation of the bulge through minor mergers in spiral 
galaxies. 

  Toth \& Ostriker (1992) have used the argument of the fragility 
of disks to constrain the frequency of merging and the amount of
accretion, and draw implications on cosmological parameters.
They claim that the thickness of the Milky Way disk
implies that no more than 4\% of its mass can have accreted
within the last $\rm 5\,10^{9}$ yrs; moreover they question the currently
fashionable theory of structure growth by hierarchical merging, which 
would not be supported by the presence of thin galactic disks, cold
enough for spiral waves to develop.

\section{Sample and observations}

Our sample consists of 24
interacting systems containing at least one edge-on galaxy.
We also observed a control sample of 7 edge-on isolated galaxies
for comparison.

Observations were carried out at the OHP 1.2 m telescope in the 
$B$, $V$ and $I$ passbands. General 
photometric results of the observations
(including isophotal maps of all objects) are presented in
Paper I.

\section{Results} 

For most galaxies, we find  a constant 
scale height with radius, within 20\%.

From the direct comparison of scale height $z_{0}$ and
$h/z_{0}$ distributions in both samples of interacting and
isolated spirals we find evidence for 
{\bf thickening} of galactic disks in
interacting systems, {\bf by a factor 1.5 to 2} (see Fig.1). 
This thickening refer to
the region of exponential disk between 1 and 2.4 of
exponential scalelength (or between 0.6 and 1.4 of
effective radius).

\begin{figure}
\centerline{\psfig{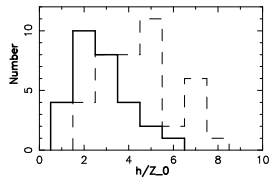}}
\caption[1]{Distribution of scalelength to scale height ratios
for interacting (thick solid line) and normal (dashed line)
galaxies. The data for normal spirals are from our observations
and literature.}
\end{figure}

The mean characteristics of edge-on interacting galaxies
in our sample are: absolute blue luminosity $M_{B}~=~\rm 
-19.6~\pm~1.0$ (so ``face-on'' magnitude must be about $\rm
-21$), exponential scalelength $h~=~\rm 4~\pm~2$ kpc
($\rm H_{0}=75\,km/s/Mpc$).
Therefore, the typical galaxy in our sample is comparable with the
Galaxy and M 31. Most edge-on galaxies in the interacting sample 
have comparable luminosity companions within one optical diameter.

One can conclude that tidal interaction
between large spiral galaxies, like the Milky Way and the
Andromeda galaxy, at a relative distance of about one optical
diameter, leads to thickening by a factor 1.5-2 of their stellar
disk.

\section{Discussion}

The $h/z_0$ ratio is 1.5-2 times smaller in interacting galaxies:
this is found to be due
not only to a higher average scale height $z_0$ in
the interacting sample, but also to a somewhat smaller scalelength $h$.

  This corresponds quite well to the predictions of N-body simulations
(Quinn et al 1993, Walker et al 1996): the gravity torques induced by
the tidal interaction produce a central mass concentration, while 
the outer disk spreads out radially, leading to a decrease of $h$.
 Most of the heating is expected to be vertical, since the planar
heating is taken away by the stripped stars either in the primary or
in the satellite. The quantitative agreement between observations
and simulations is rather good, given the large dispersion expected due 
to the
initial morphology of the interacting galaxies: a dense satellite 
will produce much more heating than a diffuse one, where stripped stars
take the orbital energy away; a mass-condensed primary will inhibit
tidally-induced spiral and bar perturbations, that are the source of 
heating both radially and vertically. 

The fact that tidal interactions and minor mergers must have concerned
every galaxy in a Hubble time, and therefore also the presently 
isolated and undisturbed galaxies, tells us that 
the lower values
of $h/z_0$ observed for the interacting sample must be transient.  
Radial gas inflow induced by the interaction may have contributed to 
{\bf reform a thin young stellar disk}, while the vertical thickening
has formed the thick disk components now observed in the Milky
Way and many nearby galaxies. 
This process might be {\bf occuring all
along the interaction}, so that the galaxy is never observed without
a thin disk. One cannot therefore date back the period of the
last interaction by the age of the thin disk, as has been proposed
by Toth \& Ostriker (1992) and Quinn et al (1993). The Milky Way,
experiencing now interactions with the Magellanic Clouds and a few
dwarf spheroidal companions, has still a substantial gaseous and 
stellar thin disk.
Further self-consistent simulations, including gas
and star-formation, must be performed to derive more significant 
predictions.

The fundamental role of the re-formation of a thin 
stellar disk is obvious in Fig.2: there are correlations
between the HI content of a galaxy and its stellar scale height
and relative thickness $h/z_{0}$. Dashed lines in Fig.2
show double regression fits for normal spirals:
$z_{0}\,\rm (kpc)\,=\,0.84\,\times\,[M(HI)/L_{B}]^{-1/2}$  and 
$h/z_{0}\,=\,\rm 5.0\,\times\,[M(HI)/L_{B}]^{1/2}$, where M(HI) is
the total HI mass (in $\rm M_{\odot}$)
and $\rm L_{B}$ is the
total luminosity of the edge-on galaxy (in 
$\rm L^{\odot}_{B}$) uncorrected for internal absorption. 
Interacting galaxies in general follow
the same relations as normal spirals but with larger scatter.
It is interesting that the Milky Way is also satisfying the above
relations. Adopting for the absolute luminosity of "edge-on" Milky Way 
$ M_{B}\rm\approx-20.5+1.5=-19.0$ and
$\rm M(HI)=8\,10^{9}\,M_{\sun}$, we obtain from the above
correlations $z_{0}~=\rm~0.74\pm0.27$ kpc and $h/z_{0}~=\rm~
5.7\pm1.7$. These values are in agreement with current
estimates of the Milky Way parameters (e.g. Sackett 1997).

\begin{figure}
\centerline{\psfig{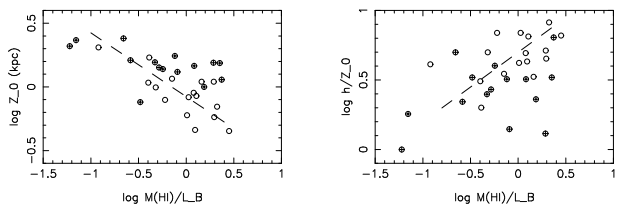}}
\caption[2]{Distribution of normal (open circles) and 
interacting 
(solid circles) galaxies in the plane 
M(HI)/$\rm L_{B}$ - $z_{0}$ (left)
and M(HI)/$\rm L_{B}$ - $h/z_{0}$ (right).}
\end{figure}

\end{document}